\documentclass[preprintnumbers, pra, floatfix,twocolumn,
preprintnumbers, letterpaper, superscriptaddress]{revtex4}
\usepackage{amsfonts}
\usepackage{amsmath}
\usepackage{amssymb,epsf}
\usepackage{latexsym}
\usepackage{graphicx,epsfig}
\usepackage{amssymb}
\usepackage{subfigure}
\usepackage[colorlinks=true,citecolor=blue,linkcolor=blue,urlcolor=black]{hyperref}
\usepackage{epstopdf}
\usepackage{color}

\def\be{\begin{equation}}
\def\ee{\end{equation}}
\def\ba{\begin{eqnarray}}
\def\ea{\end{eqnarray}}
\def\la{\langle}
\def\ra{\rangle}

\def\vep{\varepsilon}

\begin{document}
\title{Partially topological phase in a quantum loop gas model with tension and pressure}
\author{J. Abouie}
\email{jahan@iasbs.ac.ir}
\affiliation{Department of Physics, Institute for Advanced Studies in Basic Sciences (IASBS), Zanjan 45137-66731, Iran}
\author{M. H. Zarei}
\email{mzarei92@shirazu.ac.ir}
\affiliation{Department of Physics, School of Science, Shiraz University, Shiraz 71946-84795, Iran}

\begin{abstract}
Enhancing robustness of topological orders against perturbations is one of the main goals in topological quantum computing. Since the kinetic of excitations is in conflict with the robustness of topological orders, any mechanism that reduces the mobility of excitations will be in favor of robustness. A strategy in this direction is adding frustration to topological systems. In this paper we consider a frustrated toric code on a kagome lattice, and show that although increasing the strength of perturbation reduces the topological order of the system, it cannot destroy it completely. Our frustrated toric code is indeed a quantum loop gas model with string tension and pressure which their competition leads to a partially topological phase (PTP) in which the excitations are restricted to move in particular sublattices. In this phase the ground state is a product of many copies of fluctuating loop states corresponding to quasi one dimensional ladders. By defining a non-local matrix order parameter and studying the behavior of ground state global entanglement (GE), we distinguish the PTP from the standard topological phase. The partial mobility of excitations in our system is a reminiscent of fracton codes with restricted mobility, and therefore our results propose an alternative way for making such a restriction in three dimension.
\end{abstract}
\maketitle
\section{Introduction}

One of the mainstream of research in condensed matter physics and quantum computing is searching for topological phases of matter which are beyond the Landau-Ginzburg symmetry breaking paradigm\cite{t1,t2,20,Haldane,string,string2}. In topological states, the degeneracy of the ground state is protected by topological properties \cite{wen0, wen1, wen2}, and the excitations are anyonic quasi-particles displaying a fractional statistics intermediate between bosons and fermions \cite{any}. Regarding these interesting features, topological quantum systems are considered for quantum computing \cite{topo}, due to their robustness against perturbations.

Toric code is an exactly solvable topological quantum system \cite{Kitaev2003,15,pra} which is used as a robust quantum memory \cite{bravyi,del,18,19}. The ground state of the toric code is a superposition of closed-strings and is called a fluctuating loop state or a quantum loop gas \cite{string}. There are four degenerate ground states for the toric code on torus which are distinguished by different topological classes of loops and are robust against local perturbations \cite{loopgas,string3}. However, such a robustness is not unlimited in the sense that increasing perturbations finally lead to a phase transition out of the topological phase, and moving excitations in the system destroy the topological order.

Regarding the great importance of robustness of the topological quantum codes in their application as quantum memories, providing a mechanism to protect them against perturbations is an important task. Fracton codes in three dimension (3D) are an example \cite{fracton1} where mobility of excitations are limited due to quantum glassiness \cite{chamon}. Another example is a toric code with random couplings in the presence of a magnetic field where excitations are localizaed due to the disorder inherent in the code \cite{pachos}. Introducing frustration to toric code model in an external magnetic field can also lead to robustness by reducing the mobility of excitations \cite{frus1}. In this regard, it is shown that the interplay between frustration and topological order leads to a rich phase diagram in topological systems \cite{frus2}.

In this paper we consider a frustrated quantum loop gas model with tension and pressure, and show that, in a wide region in the ground state phase diagram, local perturbations reduce the topological order of the toric code, but cannot destroy it, completely. Our study is based on mappings between topological quantum codes and classical spin models \cite{zarei20,zarei18,castel2005,castel}. Recently, it has been demonstrated that the combined effects of string tension and pressure lead to a topological line with infinite robustness against perturbations \cite{zare-abo}. Our quantum loop gas model is a frustrated toric code on a kagome lattice. We investigate the ground state phase diagram of our toric code model by mapping it onto a frustrated Ising model. We show that there is a new quantum phase in the ground state phase diagram with a topological order different from the standard topological phases. In particular, we show that in this phase, dubbed "partially topological phase (PTP)", the ground state is a superposition of a partial set of possible loops. We next describe this phase in terms of the kinetic of excitations, and show that there is a partial restriction on the mobility of excitations in the PTP so that they are forced to move in particular sublattices. In order to characterize such a topological phase, we define a non-local matrix order parameter which shows the differences between this PTP and other standard topological phases. We consider a multipartite measure of entanglement which reflects highly entangled nature of topological phases. In particular, we show that global entanglement is able to identify different phases of our frustrated system as well as quantum transition points. Importantly, the value of global entanglement in the PTP confirms semi-topological nature of this phase.

The structure of this paper is as follows: In Sec.(\ref{sec1}), we introduce our idea by a simple example of a frustrated toric code model and give a qualitative argument that why we expect there is a PTP. In Sec.(\ref{sec2}), we give an exactly solvable version of the above frustrated quantum loop gas and, present the phase diagram of model. We show that in the PTP, the ground state is a product of many copies of fluctuating loop states corresponding to ladders on the lattice. In Sec(\ref{sec3}), we define a matrix order parameter for characterizing the PTP. We also show that in the PTP, there is a partial restriction on mobility of excitations. Finally, in Sec.(\ref{sec4}), we investigate the behavior of GE, and show that it can distinguish the PTP from topological and trivial phases.

\section{Quantum loop gases with tension and pressure}\label{sec1}
\begin{figure}[t]
\centering
\includegraphics[width=8cm,height=14cm,angle=0]{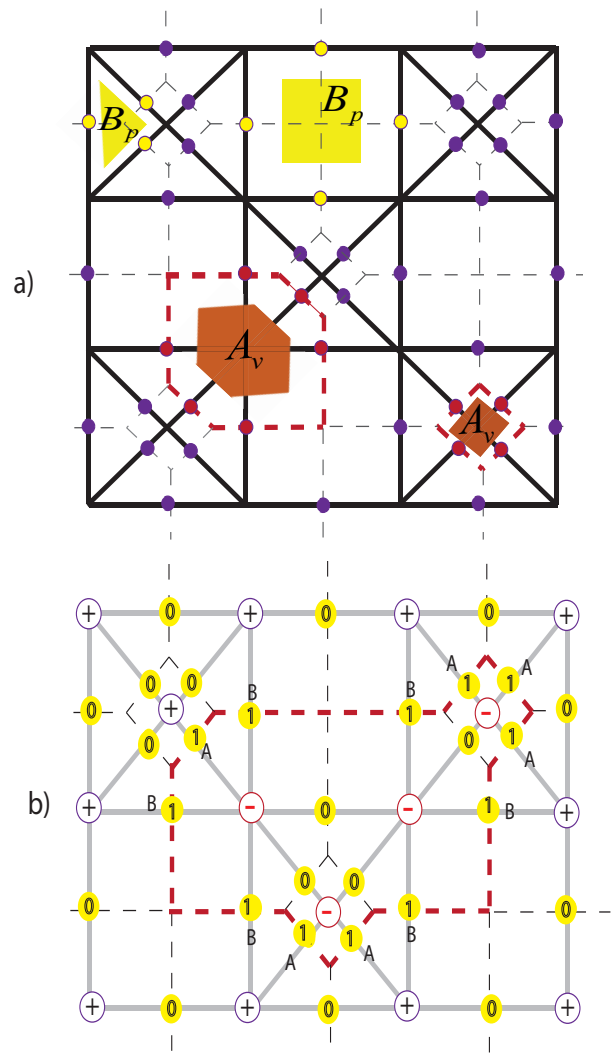}
\caption{(Color online) a) Toric code on a Kagome lattice. Qubits live on edges in the lattice. There are two types of plaquette operator $B_p$ corresponding to the triangular and square plaquettes, and two types of vertex operator which are represented by hexagonal and square loops in the dual lattice. b) Classical spins in Ising model live on vertices in the Kagome lattice. A particular configuration of spins in Ising model can be represented by a loop in the dual lattice where spins inside (outside) the loop are $-1$ ($+1$). Such a loop configuration is equivalent to a loop state in the toric code where qubits living on the loop are in the state $|1\ra$ and others are in the state $|0\ra$.  }\label{kag}
\end{figure}

A simple quantum loop gas is the toric code model defined on an arbitrary oriented two-dimensional lattice. Here, we consider a kagome lattice with qubits living on edges in the lattice. The Hamiltonian is defined in the form of
\begin{equation}
H_0 =-\sum_{p}B_p -\sum_{v} A_v,
\end{equation}
where the plaquette operator $B_p =\prod_{i\in p} Z_i$ is a tensor product of the Pauli operators $Z_i$ corresponding to the $i$th qubit around the plaquette $p$, and the vertex operator $A_v=\prod_{i\in v}X_i$ is a tensor product of the Pauli operators $X_i$ corresponding to the $i$th qubit incoming to the vertex $v$, see Fig. \ref{kag}-a. The ground state of $H_0$ is simply obtained as the operators $B_p$ and $A_v$ are commuted. Consider a product state where all qubits in the lattice are in the state $|0\ra$, the positive eigenstate of the operator $Z$. Then, as shown in Fig. \ref{kag}-b, we flip qubits corresponding to a loop configuration in the dual lattice to the state $|1\ra$, and define a loop state as $|loop\ra=\prod_{i\in loop}|1\ra_i$. Clearly $B_P |loop\ra=|loop\ra$, meaning that loop states are stabilized states of $B_p$. On the other hand, each vertex operator also corresponds to a small loop in the dual lattice. Therefore, applying a vertex operator on the state $|loop\ra$ generates another loop state corresponding to a different loop configuration. In this regard, the ground state of $H_0$ must be a superposition of all loop states i. e. a loop condensed state. In particular, if one considers each loop configuration as a non-local object, the operator $A_v$ plays the role of a kinetic term which leads to loop fluctuations and the ground state is simply a quantum loop gas.

The above quantum loop gas has a topological ground state with long range entanglement \cite{chen}. By putting the gas under a string tension, by adding a Zeeman term to the initial Hamiltonian $H_0$ in the form of $-h\sum_i Z_i$ with $h> 0$ being an external magnetic field, smaller loops appear that reduce the topological order of the gas. Increasing the string tension removes loop superpositions and destroys the topological order completely at a critical field $h*=0.33$.
In a similar way, in the presence of a magnetic field in opposite direction ($h<0$), the field behaves as a pressure in the loop gas. Although pressure is opposite of tension and leads to the creation of larger loops, but similarly, it destroys the topological order of the gas, too.

Regarding the above mentioned opposite roles of the string tension and pressure, it is interesting to study their combined effects. To this end, we consider a different Hamiltonian given by:
\begin{equation}\label{tr}
H=H_0 - h_A \sum_{i\in A} Z_i -  h_B \sum_{i\in B} Z_i,
\end{equation}
where $A$ and $B$ contain qubits living on diagonal and vertical-horizontal edges in the kagome lattice, respectively. Here, $h_A $ and $h_B$ are external magnetic fields applied to qubits of type $A$ and $B$, respectively. As shown in Fig. \ref{kag}-b, each loop configuration in the lattice crosses both types of qubit. In this regard, the Zeeman terms induce an energy shift of $\Delta\vep=\Delta\vep_A+\Delta\vep_B$ with $\Delta\vep_A=(2L_A -N_A) h_A$ and $\Delta\vep_B=(2L_B -N_B) h_B$, where $N_A$ ($N_B$) is total number of $A$ ($B$) qubits and $L_A$ ($L_B$) is number of $A$ ($B$) qubits crossed by the corresponding loop configuration. Therefore, the value as well as the sign of $h_A$ and $h_B$ are crucial in determining the lowest energy state of the Hamiltonian (\ref{tr}).
In particular, there are three different possibilities: 1) In the presence of very weak magnetic fields, i.e. when $h_A, h_B\ll 1$, the energy shift is $\Delta\vep\sim 0$ and there is no considerable difference between different loop states, and hence the topological order of $H_0$ persists. 2) If one of the fields is very strong, i.e. either $h_A\gg 1$ or $h_B\gg 1$, the energy shift becomes either $\Delta\vep_A$ or $\Delta\vep_B$. In this case the system will choose small or large loops regarding the sign of the strong field, and consequently there will occur a transition to a topologically trivial phase.
3) There is an intermediate regime in which both $h_A$ and $h_B$ are $\sim 1$, but have different signs. In this case, the energy shift is $\Delta\vep\sim 2(L_A-L_B)+N_B-N_A$, which can not be simply analyzed for different loop states. In this case our quantum loop gas is actually under both the pressure and tension, simultaneously and a frustration arises in the ground state of the system.  One of the possible scenarios we can consider is the emergence of an interesting phase with a partially topological order. In particular, consider a partial set of loop states $\Omega$ including both small and large loop configurations so that the energy shift $\Delta\vep$ is the same for all members of $\Omega$. Therefore, the ground state would be a superposition of all loop states belonging to $\Omega$. Such a partially loop-condensed state is a topological phase, distinct from the topological and trivial phases of toric codes.

\section{PTP in an exactly solvable model}\label{sec2}
The above qualitative argument shows that a frustrated combination of string tension and pressure in quantum loop gas model leads to a PTP. However, in order to realize such a possibility, explicitly, we should find the whole phase diagram of the model. The Hamiltonian in Eq. (\ref{tr}) is not exactly solvable and finding its ground state for arbitrary strengths of $h_A$ and $h_B$ is not straightforward. On the other hand, adding the Zeeman term to the toric code Hamiltonian is not the only way for considering frustration effects of the string tension and pressure. In \cite{zare-abo}, we have shown that the same effects can be realized by adding some particular non-linear interactions to the toric code Hamiltonian. Here, as \cite{zare-abo}, we consider the toric code on the kagome lattice in presence of a non-linear interaction of the following form:
\begin{equation}
\sum_v e^{-\beta [h_A\sum_{i\in A}Z_i +h_B\sum_{i\in B}Z_i] },
\label{Eq:pert}
\end{equation}
where $\beta$ is an additional coupling parameter. For very small values of $\beta$, the exponential in (\ref{Eq:pert}) reduces to $1-\beta [h_A\sum_{i\in A}Z_i +h_B\sum_{i\in B}Z_i]$, which is the standard Zeeman term. The ground state of the toric code in the presence of the above mentioned term can be exactly obtained in the following form \cite{zarei20}:
\begin{equation}\label{fg}
|G(\beta)\ra= \frac{1}{\sqrt{\mathcal{Z}}}e^{\frac{\beta}{2}[h_A\sum_{i\in A}Z_i +h_B\sum_{i\in B}Z_i]}|G_0\ra,
\end{equation}
where $|G_0\ra$ is the ground state of the bare toric code and $\mathcal{Z}$ is a normalization factor. Now, it is important to show how the non-linear interactions in (\ref{Eq:pert}) play exactly the role of the string tension or pressure similar to what we observed for the Hamiltonian (\ref{tr}). In the ground state $|G(\beta)\ra$ the amplitude of each loop configuration depends on $h_A L_A$ and $h_B L_B$ as $e^{\frac{\beta}{2}\Delta\vep}$. This clearly implies that increasing $\beta$ decreases the amplitudes in favor of smaller or larger loops depending on the value and the sign of $h_A$ and $h_B$. These are all things we need for the frustration effect we discussed in previous section.

\subsection{Mapping onto a frustrated Ising model}

Regarding the special form of the ground state in Eq. (\ref{fg}), the normalization factor $\mathcal{Z}$ is equal to the partition function of a two dimensional Ising model defined on the kagome lattice with classical spins residing at the vertices. The Hamiltonian of such a dual Ising model is given by
\begin{equation}
{\mathcal H}=-h_A\sum_{\la \mu,\nu \ra\in A}S_{\mu} S_{\nu} -h_B\sum_{\la \mu,\nu \ra\in B}S_{\mu} S_{\nu},
\label{Eq:Ising-Hamiltonian}
\end{equation}
where $S_\mu$ is the classical spin resided on the site $\mu$, and the sums run over nearest neighbors in the sublattices $A$ and $B$. The coupling parameter $\beta$ in the quantum model corresponds to $1/k_{\rm B} T$ in the classical Ising model where $k_B$ is the Boltzmann constant and $T$ is temperature.
In order to prove the above mapping, we should notice that, similar to the $|G_0 \rangle$, $|G(\beta)\rangle$ is also a superposition of all loop states with the difference that each loop state has a particular amplitude. To see this, consider a loop state corresponding to a particular loop configuration which crosses $L_A$($L_B$) numbers of qubits belonging to the edges of $A$ ($B$). When we apply $e^{-\frac{\beta}{2}[h_A\sum_{i\in A}Z_i +h_B\sum_{i\in B}Z_i]}$ on such a loop state, a term with amplitude $e^{\frac{\beta}{2}\Delta\vep}$ will be generated. This amplitude is equal to the square root of the Boltzmann weight of corresponding spin configuration in the Ising model. To show this, corresponding to a loop configuration we set the classical spins at vertices inside (outside) loops to $-1$ ($+1$), see Fig. \ref{kag}-b. The energy of such a spin configuration in the Ising model is equal to $\Delta\vep$. Therefore, $e^{\frac{\beta}{2}\Delta\vep}$ will be the square root of the Boltzmann weight of this configuration. In this regard, the normalization factor in Eq (\ref{fg}) is equal to summation of all Boltzmann weights which is the same as the partition function of the Ising model. We can summarize the above mapping as a mapping from the interaction term $S_\mu S_\nu$ in the Ising model onto the Pauli operator $Z_i$ in the ground state $|G(\beta)\ra$ where the $i$th qubit lives on the edge with two end-points at $\mu$ and $\nu$.

\begin{figure}[t]
\centering
\includegraphics[width=8cm,height=8cm,angle=0]{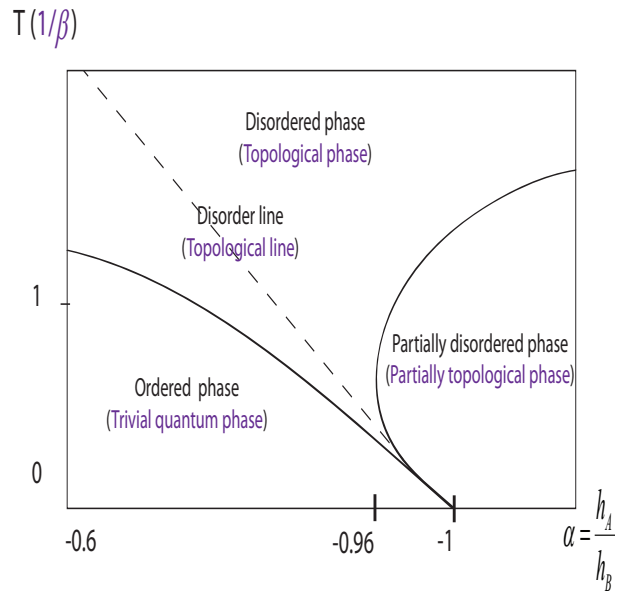}
\caption{(Color online) Phase diagram of the frustrated Ising model (\ref{Eq:Ising-Hamiltonian}) on the Kagome lattice. There are three different classical phases including disordered, ordered and partially disordered phases. Corresponding to different classical phases, there are three quantum phases in our frustrated quantum loop gas model. Disordered (ordered) phase in the classical model is equivalent to topological (trivial) phase, and corresponding to the partially disordered phase there is a PTP. Vertical axis in the classical model is temperature which is mapped onto $1/\beta$ in our quantum loop gas model. Solid lines are phases border and the dashed line in the topological line at which the topological order survives even in the presence of strong perturbations}\label{part}
\end{figure}

\begin{figure}[t]
\centering
\includegraphics[width=8cm,height=8cm,angle=0]{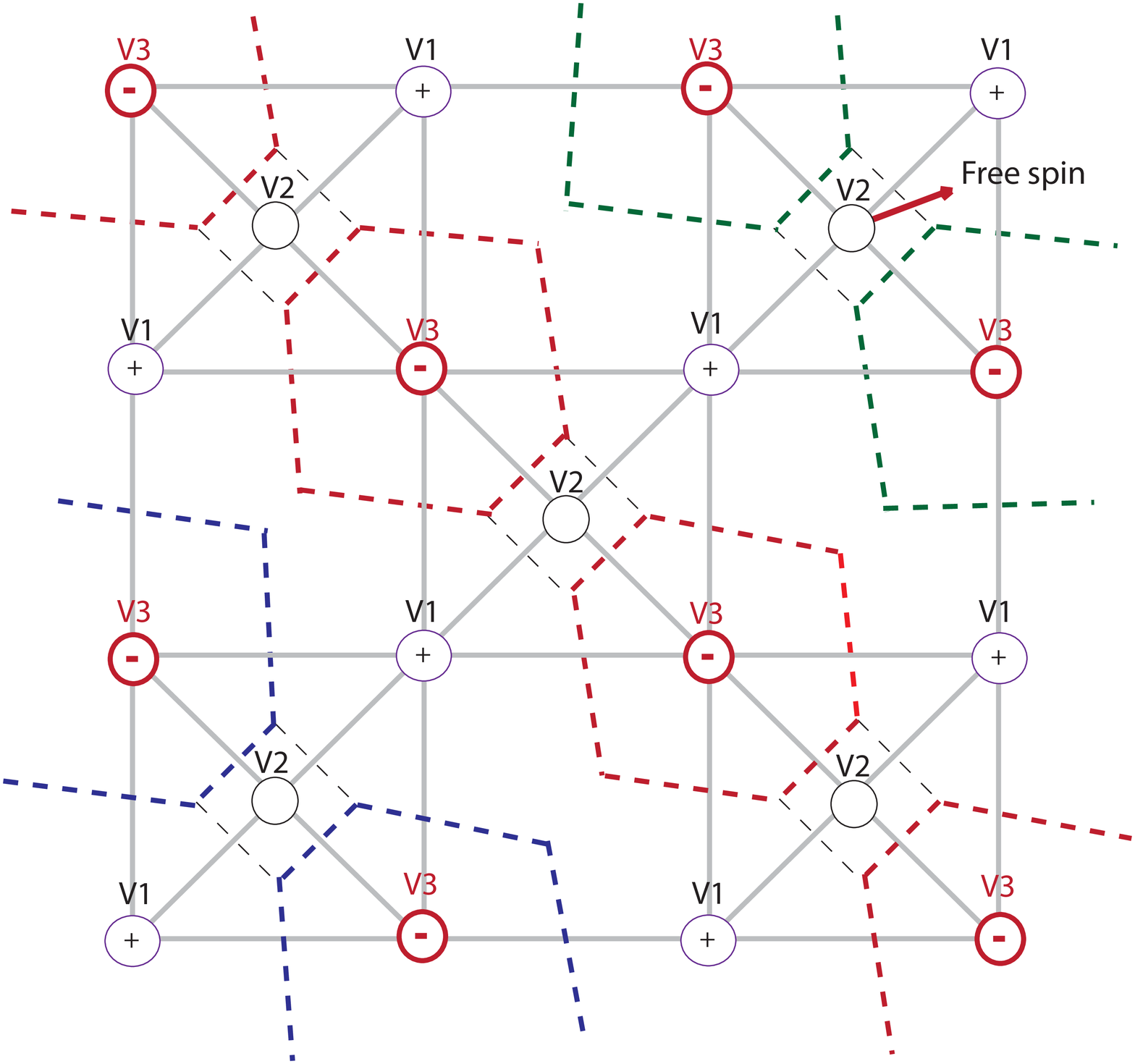}
\caption{(Color online) A loop configuration in the PTP at $\beta=\infty$. The sublattices $V1$, $V2$ and $V3$ contain the classical spins which interact with each other with the Ising Hamiltonian (\ref{Eq:Ising-Hamiltonian}). At $T=0$, in the partially disordered phase, the spins in the sublattice $V1$ ($V3$) are $+1$ ($-1$), and those in the sublattice $V2$ are completely free to be $+1$ or $-1$. Such a spin configuration in the Ising model corresponds to particular loop configurations in our quantum loop gas model. According to the sign of the spins in the sublattice $V2$, loop configurations form in diagonal direction, and the ground state $|G(\beta=\infty)\ra$ is a product of loop states fluctuating along ladder-like subspaces.}\label{partial}
\end{figure}

The above mapping can be used for characterizing the phase diagram of our frustrated toric code model. Fortunately, the frustrated Ising model on kagome lattice is well-studied \cite{diep,Steph1}, and as shown in Fig. \ref{part} there are three different phases in the phase diagram of the system. Interestingly, besides ferromagnetic and paramagnetic phases there is also a partially disordered phase in which the
spins on one sublattice are ordered magnetically, whereas on other one they are free to be +1 or $-1$.
This interesting phase diagram can be characterized by sublattice order parameters. To this end, consider three different sublattices for the kagome lattice: $V1$, $V2$ and $V3$ (see Fig. \ref{partial}). The ensemble averages of spins in these sublattices are given by the magnetizations $m_1$, $m_2$ and $m_3$, respectively. In the disordered phase (ordered phase) the above three order parameters are zero (non-zero). However, in a partially disordered phase the order parameter $m_2$ is zero while $m_1$ and $m_3$ are non-zero. This indicates that in this phase the spins in the sublattice $V2$ possess no magnetic order.

\subsection{Ground state phase diagram}

Now, we use the results of the Ising model and present the ground state phase diagram of our frustrated quantum loop gas model. All phase transitions in the frustrated Ising model correspond to singularities in thermodynamic functions like heat capacity. Therefore, it is crucial to find the equivalence of such thermodynamic functions in our quantum loop gas model.
As we discussed, since Boltzmann weights in the Ising model are encoded in the ground state of the quantum loop gas, one can simply find a mapping between the ground state fidelity in our quantum model and the heat capacity in the Ising model \cite{zarei20,zanardi}. The ground state fidelity is simply defined as $\la G(\beta)|G(\beta +d\beta)\ra$ which can be approximated as:
\begin{equation}
F=1-g_{\beta \beta}d\beta^2,
\end{equation}
where $g_{\beta \beta}=\frac{\partial^2 F}{\partial \beta^2}$ is the fidelity susceptibility. It has been shown that $g_{\beta\beta}$ is related to the heat capacity of the Ising model as $g_{\beta\beta}=C_v/8\beta^2$. Accordingly, it is clear that any singularity in the heat capacity of the frustrated Ising model, at a critical temperature $T^*$, corresponds exactly to a singularity in the fidelity susceptibility in the frustrated quantum loop gas, at a critical $\beta^*=1/k_{\rm B}T^*$.

Moreover, there is a one-to-one correspondence between classical phases in the frustrated Ising model and quantum phases in our frustrated quantum loop gas, see Fig. \ref{part}. Since $\beta$ in the frustrated quantum loop gas is equivalent to temperature inverse in the Ising model, it is concluded that the paramagnetic (ferromagnetic) phase at high (low) temperatures is equivalent to the topological (trivial) phase at small (large) values of $\beta$ \cite{zarei20}. There is also a partially disordered phase in the frustrated Ising model which is equivalent to a topological phase different from the standard topological order of the toric code.  As seen in Fig. \ref{part}, for large values of $1/\beta$, and any value of $\alpha$ the system is in the topological phase. By decreasing $1/\beta$ (increasing the perturbation strength) perturbation tries to destroy the topological order. For $\alpha<0.96$ it is completely destroyed at a critical point where a topological-trivial phase transition occurs in the system, but for $\alpha>1$ although the perturbation imposes some changes, the topological order survives at least partially. In this region, by decreasing $1/\beta$ a phase transition to the PTP happens at a critical point. The interval $0.96<\alpha<1$ is interesting. In this region interestingly a reentrant phase transition occurs in the system, i.e. by increasing perturbation the system goes to the PTP, and by more increasing perturbation although we expect the topological order to be disappeared completely, but the system prefers to go back to the topological phase. This reentrance phenomenon in our quantum loop gas is due to the frustration and can be seen also in different spin models \cite{fr}. The dashed line in the topological phase is the topological line introduced in \cite{zare-abo}. At this line the topological order is robust against the nonlinear perturbation of type (\ref{Eq:pert}) with arbitrary strength, and no phase transition happens in the system by increasing $\beta$. This phenomenon is a combined effect of frustration and the nonlinearity existing in our quantum loop gas model.

\section{Characterizing the PTP}\label{sec3}

According to the mapping between the amplitudes in the ground state of our quantum model and the Boltzmann weights in the Ising model, the ground state of our quantum model in the PTP should be a superposition of loop states corresponding to spin configurations in the partially disordered phase in the Ising model.
In order to obtain the allowed loop configurations we look at the sign of the classical spins in the Ising model and draw a loop around the spins $-1$.
In the partially disordered phase, at $T=0$, all spins in the sublattice $V1$ are $+1$ and therefore they should not be surrounded by any loop, while all spins in $V3$ are $-1$ and they must be surrounded by loops. But the spins in $V2$ behave differently. Due to the frustration, they are free to be $+1$ or $-1$, and therefore their sign determines their status. In this regard, in the PTP of our frustrated quantum model, in the limit of infinite $\beta$, we have a particular set of loop configurations. In this phase loop configurations are constructed by small loops encompassing the spins $-1$ in sublattice $V2$ and all spins in $V3$. Therefore, the partially topological state is a loop condensed state written as a superposition of particular loop configurations as:
\begin{equation}\label{rty}
|G_{\beta=\infty} \ra = \prod_{v\in V2} (1+A_v)\prod_{v\in V3}A_v|0\dots0\ra,
\end{equation}
where we ignored the normalization factor. Since each $A_v$ corresponds to a small loop around the vertex $v$, applying $\prod_{v\in V3}A_v$ on $|0\dots0\ra$ results in a particular loop state including small loops around vertices belonging to $V3$. On the other hand, $\prod_{v\in V2} (1+A_v)$ is also a superposition of different loop configurations generated by loops around vertices belonging to $V2$. In this regard, the state $|G_{\infty}\ra$ is simply a superposition of loop configurations constructed by $V2$ and $V3$-type loops. In particular, notice that since there is no loop around vertices belonging to $V1$, the state $|G_{\infty}\ra$ is in fact a product of quantum states corresponding to diagonal ladders, as shown in Fig. \ref{partial}, where each ladder corresponds to a fluctuating loop state constructed by $V2$-type and $V3$-type loops. In other words, while the ground state at $1/\beta =0$ is still a loop state, fluctuating along one-dimensional ladders.
In order to show that the PTP has topological features, following we define a nonlocal order parameter to distinguish it from the standard topological phases.

\subsection{Matrix order parameter}
\begin{figure}[t]
\centering
\includegraphics[width=7cm,height=5cm,angle=0]{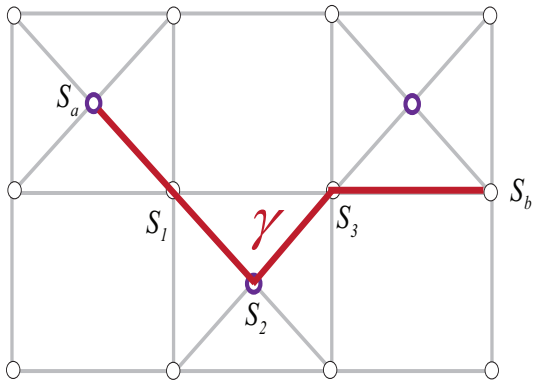}
\caption{(Color online) A schematic illustration of a string $\gamma$ which starts from a vertex $a$, passes through different classical spins and terminates at a vertex $b$. Corresponding to the string $\gamma$ there is a string operator in our quantum loop gas model which is equal to the product of $Z$ operators living in the middle of the links that connect neighboring classical spins.}\label{order}
\end{figure}

\begin{figure}[t]
\centering
\includegraphics[width=8cm,height=6cm,angle=0]{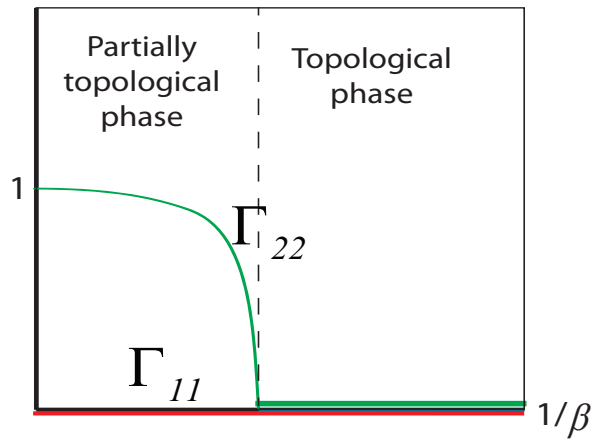}
\caption{(Color online) A schematic illustration of the string order parameters versus $1/\beta$, for $|\alpha|>1$. In the topological phase both the order parameters $\Gamma_{11}$ and $\Gamma_{22}$ are zero. By decreasing $1/\beta$ the order parameter $\Gamma_{22}$ starts to increase at a critical point, and a phase transition from the topological phase to the PTP occurs in the system. In the PTP while $\Gamma_{11}=0$, $\Gamma_{22}$ becomes non-zero which indicates the loss of part of topological order. }\label{m-order}
\end{figure}

Here, we define a non-local order parameter to characterize the PTP, seen in our frustrated quantum loop gas. As mentioned the connection between the frustrated Ising model on a kagome lattice and our frustrated quantum loop gas is simply obtained by mapping the interaction term $S_\mu S_\nu$ in the classical model onto the Pauli operator $Z_i$ in the quantum model, where the site $i$ is located in the middle of the link $\mu\nu$ (see Fig. \ref{kag}). In this mapping the thermal expectation value $\la S_\mu S_\nu \ra$ is equivalent to the quantum expectation value $\la Z_i \ra=\la G(\beta)|Z_i |G(\beta)\ra$. Moreover, we are also able to obtain spin-spin correlation functions. As shown in Fig. \ref{order}, the two spins $a$ and $b$ can be connected to each other by an open string $\gamma$. Since for each spin variable we have $S^2 =1$, we can rewrite the spin-spin interaction $S_a S_b$ as $(S_a S_1)(S_1 S_2)(S_2 S_3)\dots(S_N S_b)$ where $S_1, S_2,\dots, S_N$ are spins that the string $\gamma$ crosses.
Next, according to our quantum-classical mapping, the ensemble average of $(S_a S_1)(S_1 S_2)(S_2 S_3)...(S_N S_b)$ is equivalent to the expectation value of $Z_i$ operators on the quantum ground state:
\begin{equation}
\la (S_a S_1)(S_1 S_2)(S_2 S_3)\dots(S_N S_b)\ra_{en.} =\la \prod_{i\in \gamma} Z_i \ra_{g.s.}=\Gamma_{ab},
\end{equation}
where $i \in \gamma$ refers to all qubits living on edges that make up the string $\gamma$. This means that each spin-spin correlation function in the classical model is equivalent to a string operator in our quantum model where classical spins are living at the two end-points of that string. If the distance between spins goes to infinity, the spin-spin correlation in the Ising model will be simply related to the order parameter in the form of:
\begin{equation}\label{wer}
\Gamma_{ab}=\la S_a S_b \ra_{en.}=\la S_a \ra \la S_b \ra =m_a m_b,
\end{equation}
where $m_a$ and $m_b$ are magnetizations in the Ising model. Since the distance between $a$ and $b$ goes to infinity, the corresponding string operator in the quantum model, with endpoints at $a$ and $b$, can be regarded as a non-local order parameter for characterizing topological phase transitions. According to Eq. (\ref{wer}), we have nine string operators $\Gamma_{ij}$ with $i,j=1, 2$ and 3. However, since $m_1$ and $m_3$ have the same value but different signs, we can capture the entire ground state phase diagram by only four parameters $\Gamma_{ij}$ with $i, j=1$ and 2, where form a $2\times 2$ matrix order parameter:
\begin{equation}
\Gamma=\left(
  \begin{array}{cc}
    \Gamma_{11} & \Gamma_{12} \\
    \Gamma_{21} & \Gamma_{22} \\
  \end{array}
\right),
\end{equation}
with $\Gamma_{12}=\Gamma_{21}=\sqrt{\Gamma_{11}\Gamma_{22}}$. In Fig. \ref{m-order}, we have illustrated a schematic plot of $\Gamma_{11}$ and $\Gamma_{22}$ versus $1/\beta$ for a given $|\alpha | > 1$. As seen, for large values of $1/\beta$, when the system is in the topological phase, both these parameters are zero. By decreasing $1/\beta$, $\Gamma_{22}$ starts to increase at a critical point where the system enters to the PTP. In this phase $\Gamma_{22}$ increases by decreasing $1/\beta$ but $\Gamma_{11}$ remains zero. By more decreasing $1/\beta$ both the parameters $\Gamma_{11}$ and $\Gamma_{22}$ become nonzero (not shown) where a phase transition to the trivial phase occurs.

\subsection{Excitations}
For better understanding the nature of our PTP, it is useful to present the ground state of our frustrated quantum loop gas in a basis constructed by the eigenstates of the Pauli operator $X$. Actually the perturbation (\ref{Eq:pert}) is in terms of $Z$ operators, and therefore examining the behavior of excitations generated by $Z$ operators is worthwhile for understanding the topological features of this phase. To this end, we consider the ground state $|G_{\beta=\infty}\ra$ which is stabilized by the operators $A_v$ and $B_p$ where $v \in V2$ and $p$ runs over all plaquettes. If $M$ is the number of all palquettes and vertices, the number of vertices in $V2$ in addition to the number of all plaquettes is equal to $M-(N_1+N_3)$, with $N_1$ and $N_3$ being the number of vertices belonging to the sublattices $V1$ and $V3$, respectively.
Therefore, for characterizing $|G_{\beta=\infty} \ra$ as a stabilizer state, we need $N_1+N_3$ independent stabilizer operators with the eigenstate $|G_{\beta=\infty}\ra$ and the eigenvalue $+1$.
Considering that all the $N_1$ independent string operators $\Gamma_{11}$,
and all the $N_3$ independent string operators $\Gamma_{33}$
commute with $A_{v\in V2}$, due to having an even number of shared qubits,
they stabilize the state $|G_{\beta=\infty} \ra$.
The ground state $|G_{\beta=\infty} \ra$ can be expressed in terms of the above complete set of stabilizers as:
\begin{equation}\label{eqp}
|G_{\beta=\infty} \ra = \prod_{p} (1+B_p)\prod_{\gamma_{11}}(1+\Gamma_{11})\prod_{\gamma_{33}}(1+\Gamma_{33})|+\dots+\ra,
\end{equation}
where the product $\prod_{\gamma_{11(33)}}$ runs over $N_1$ ($N_3$) independent string operators, and $|+ \ra= \frac{1}{\sqrt{2}}(|0\ra + |1\ra)$ is the eigenstate of the Pauli operator $X$ with eigenvalue $+1$.  Now, we compare the state $|G_{\beta=\infty} \ra$ with the ground state of the pure toric code $|G_0 \ra =\prod_p (1+B_p) |+\dots+\ra$. Since $B_p$ is a loop operator in the Kagome lattice, and $Z|+\ra =|-\ra$, $|G_0\ra$ is a superposition of all loop configurations constructed by $|-\ra$s in a sea of $|+\ra$s. However, the state $|G_{\beta=\infty}\ra$ in (\ref{eqp}) has an important difference from $|G_0\ra$; that is,  in addition to the loop operator $B_p$ there are also open strings $\Gamma_{11}$ and $\Gamma_{33}$. Therefore, the ground state in the PTP is a superposition of all configurations formed by loops and open strings with end-points living in the vertices belonging to $V1$ and $V3$.
It is interesting to interpret this result from the viewpoint of moving excitations. Since the perturbation term in our quantum loop gas is constructed of $Z$ operators, it leads to open strings that tries to remove the topological order of the ground state. However, our result shows that due to the frustration there are three types of excitations in the PTP: excitations of type-1, 2, and 3 living in the $V1$, $V2$ and $V3$ sublattices, respectively. In the PTP, the system is robust against excitations of type-2, but the excitations of type-1 and 3 can move in the corresponding sublattices, and reduce the topological order of the system. This restriction on the mobility of excitations is an important property of the PTP that can be compared with other topological systems like fracton codes on 3D lattices. Fracton excitations are restricted to move in certain subspaces due to quantum glassiness \cite{chamon}, and as a result a topological overprotection occurs in fracton codes. In our quantum loop gas model, frustration creates string tension and pressure which lead to such an overprotection. Although frustration does not freeze excitations but limits their mobility in ladder-like subspaces.

\section{Global entanglement}\label{sec4}

Topological orders can also be characterized by entanglement which reveals the non-local nature of topological phases \cite{Kitaev06, Li08, Polmann10, Ahmadi20, Ahmadi22, Fidkowski10}.
A multipartite measure of entanglement is global entanglement which is defined as $GE=2[1-\frac{1}{N}\sum_i tr(\rho_i ^2)]$, where $\rho_i$ refers to the reduced density matrix corresponding to the $i$th qubit. It has been shown that the global entanglement in the toric code model in the presence of the nonlinear perturbation (\ref{Eq:pert}) is equivalent to the internal energy in the Ising model \cite{elahe}.  Actually the linear entropy for an edge qubit $i$ in the perturbed toric code model is equal to the interaction term $S_{\mu} S_{\nu}$ in the classical Ising model, where $\mu$ and $\nu$ are two classical spins resided on the two end-points of the edge $i$. Therefore, the global entanglement in our frustrated quantum loop gas is given by:
\begin{equation}
GE=1-\frac{1}{2}(\la S_{\mu} S_{\nu} \ra_A^2 +\la S_{\mu} S_{\nu} \ra_B ^2),
\end{equation}
\begin{figure}[t]
\centering
\includegraphics[width=8cm,height=12cm,angle=0]{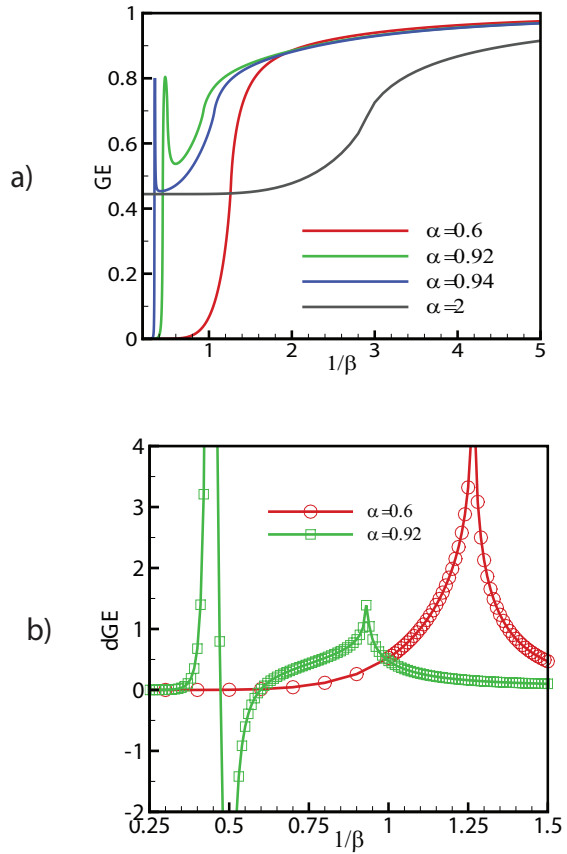}
\caption{(Color online) a) Global entanglement versus $1/\beta$, for different values of $\alpha$. For $|\alpha|<0.96$ there is a phase transition from trivial to topological phase, and the GE changes from zero in the trivial phase to 1 in the topological phase.  For $0.96<|\alpha| <1$ three phase transitions happen by increasing $1/\beta$. At the first transition point, GE changes from zero in the trivial phase to near 1 in the topological phase, at the second transition point GE changes from 1 to near $1/2$ in the PTP, and finally at the third transition point GE changes from $1/2$ to $1$ in the topological phase. b) Derivative of $GE$ versus $1/\beta$. The cusps and infinite picks reveal exactly the location of phase transitions.}\label{global}
\end{figure}
where $\la S_{\mu} S_{\nu}\ra_{A (B)}$ refers to a link energy in $A (B)$. Using the free energy of the frustrated Ising model (see \cite{diep}), one can readily obtain these expectation values and consequently the $GE$ of our quantum loop gas. In Fig. \ref{global}-a we have plotted the $GE$ versus $1/\beta$ for various $\alpha$. According to the phase diagram of our frustrated quantum loop gas, shown in Fig. \ref{part}, in the narrow interval of $0.96< \alpha <1$ where the reentrance phenomenon can occur, the system experiences different phase transitions by increasing $\beta$.
Interestingly, the different phase transitions are completely observed in the behavior of the $GE$. As seen, for large enough $1/\beta$ where the system is deeply in the topological phase, the $GE$ has its maximum value, i.e. $GE=1$. Far from the reentrance phenomenon, for $\alpha>1$ ($\alpha<0.96$) by decreasing $1/\beta$, $GE$ decreases monotonically, falls down to $0.5 (0)$ at the topological-partial topological (topological-trivial) phase transition point, and sticks to $0.5~(0)$ in the partial topological (trivial) phase.  In the reentrance interval (see $\alpha =0.92,~0.94$) $GE$ behaves oscillatory, it is $1$ in the topological phase, tries to touch $0.5$ in the PTP, and becomes $0$ in the trivial phase. The transition points are locations that the $GE$ curve concavity changes. These changes are reflected as cusps and divergence in the $GE$ derivative, as clearly seen in Fig.  \ref{global}-b.

\section{Summary and outlook}
It is known that kinetic of excitations in a topological phase is strongly reduced due to frustration. Here, we examined the ground state phase diagram of
a frustrated quantum loop gas model, and demonstrated that there is a PTP in which kinetic of excitations is partially reduced in the sense that there is a restriction for mobility of excitations living in some sublattices. In this phase the ground state is a tensor product of quasi one dimensional fluctuating loop states, there are different types of excitation living in three different sublattices, and excitations belonging to a particular sublattice can move freely. Frustration is a key factor to realizing this phase where string tension and pressure lead to the PTP.
This phase can in principle be seen in frustrated toric code in the presence of Zeeman fields, but we considered an exactly solvable model which contains beyond the Zeeman terms.
To obtain the ground state phase diagram, we mapped our frustrated quantum loop gas model onto a classical Ising model. To characterize this phase, we defined a matrix order parameter and also considered global entanglement, and showed that these quantities can well distinguish the PTP from other phases of the system.

Topological excitations with restricted mobility are also seen in three dimensional fracton codes due to quantum glassiness. In this regard, our results propose that similar features can also be realized in two dimension due to frustration. Moreover, randomness can also lead to the localization of excitations, so it is interesting to study the combined effects of geometrical frustration and randomness in toric codes, and check the possibility of having topological excitations with restricted mobility in certain one-dimensional sublattices similar to lineons in fracton codes.

\section*{Acknowledgement}
J. A. is grateful to the Institute for Advanced Studies in Basic Sciences, for financial support through research grant No. G2021IASBS12654.

\end{document}